\documentclass[superscriptaddress,twocolumn,nofootinbib]{revtex4}
\usepackage{amssymb}
\usepackage{amsfonts}
\usepackage{amsmath}
\usepackage{graphicx}
\usepackage{dcolumn}
\usepackage{bm}
\usepackage[latin1]{inputenc}
\usepackage[dvips]{hyperref}

\begin{document}

\title{Primordial brusque bounce in Born-Infeld determinantal gravity}

\author{Franco Fiorini}
\email[Member of Carrera del Investigador Cient\'{\i}fico (CONICET,
Argentina); ]{francof@cab.cnea.gov.ar}
\affiliation{Instituto Balseiro, Universidad Nacional de Cuyo, R8402AGP Bariloche, Argentina.}


\begin{abstract}
We study a particular exact solution to the Born-Infeld determinantal gravity consisting of a cosmological model which undergoes a \emph{brusque bounce}. The latter consists of an event characterized by a non-null (but finite) value of the squared Hubble rate $H^{2}$ occurring at a minimum (non-null) scale factor. The energy density and pressure of the fluid covering the whole manifold are perfectly well behaved in such an event, but the curvature invariants turn out to be undefined there because of the undefined character of $\dot{H}$. It is shown that the spacetime results geodesically complete and singularity free, and that it corresponds to a picture of an eternal Universe in which a (somewhat unconventional) bounce replaces the standard Big Bang singularity. This example tends to emphasize that, beyond Einstein's theory of General Relativity, and in the context of extended theories of gravity formulated by purely torsional means, the criterion of a singularity based on pathologies of scalars constructed upon the Riemann curvature tensor, becomes objectionable.
\end{abstract}

\maketitle

\section{Introduction}\label{intro}

Since the early days of the new phase of the General Theory of Relativity (GR), crowded by the singularity theorems of Hawking and Penrose \cite{H-P}, \cite{H-E}, the very concept of singularity was the subject of great concern for the mathematical relativist. The root of this issue was mastered discussed in the pioneer Galilean dialogue between Sagredo and Salviati elaborated in Ref. \cite{Geroch}. What the latter was trying to explain to the former (who inevitably brought up what he thought was a good electromagnetic analogy), was that the problem in defining a spacetime singularity relies on the existence of points that,--most of the time--are not necessarily part of the spacetime structure itself. This harmless and almost semantic property raised enormous difficulties at the time of being concrete regarding what a singularity is, for its mere existence should be inferred from the points in the manifold close to the conflictive \emph{point} (in some appropriated mathematical sense).

 One hundred years after its conception, there exists now a wide repertoire of ideas on what can be considered to be a singular state within the context of GR\footnote{For an early classification, see Refs. \cite{Clarke1}-\cite{TCE}} \cite{Joshi1}. Actually, it is remarkably easy to produce some sort of singular state by cutting out and/or excising subsets of a given (otherwise regular) spacetime, even though the manifold so obtained might have a definitely dubious physical interpretation. For instance, by removing a point from Minkowski spacetime we can clearly get geodesics that abruptly end at such a removed point, indicating that the observers represented by these curves will impedingly cease to exist there. Even though this example seems deliberately artificial, it teaches us an important lesson. In order to arrive to the concept of singularity, and irrespective of the potential bad behavior of the metric tensor and its related curvature in or near the conflictive point, what turns out to be relevant is the behavior of causal curves and geodesics in the spacetime. As a matter of fact, the key idea of what now is widely accepted as the most fruitful condition for a spacetime to be defined as singular
is that it be \emph{timelike} (\emph{null}) \emph{geodesically incomplete}, i.e., if it contains a maximally extended timelike (null) geodesic whose affine parameter does not assume values in the full range from $-\infty$ to $+\infty$.

Even this quite accepted definition has its deficiencies. On the one hand, there exist examples of geodesically complete spacetimes that admit incomplete (non-geodesic) curves. Although this is regarded as unimportant in the context of the usual spatially homogeneous and isotropic cosmological FRW models, it is certainly relevant in more generic spacetimes. In those cases, the endpoint of the incomplete curves in $\mathcal{M}$ is incorporated as a regular point in the boundary $\partial$ of an extended manifold $\tilde{\mathcal{M}}=\partial\cup\mathcal{M}$, called the $b$-completion of $\mathcal{M}$ \cite{Schmidt1}-\cite{Schmidt2} (a generalization of this seminal construction may be found in \cite{Szekeres}). On the other hand, more modern developments show that other kinds of singularities are not based on geodesic completeness. This is the case of the so called sudden singularities \cite{Barrow1}-\cite{Barrow4}, where the pressure of the matter fields becomes divergent in a late event whose fatality, however, is unseen by the geodesics \cite{Lazkoz}.

In any case, it is clear that the singularities in all their facets are an essential and inextricable part of the conceptual body of GR, for a large class of solutions of Einstein's field equations are singular. One might regard this abundance of singular spacetimes as a reminder that GR has only a finite range of applicability. Although the theory has successfully passed an important number of experimental tests over a wide spectrum of scales, including a couple concerning the emission of gravitational radiation \cite{Taylor}, \cite{Ligo}, it is clear that further experimental evidence in the strong gravitational field regime, where $GM/R\, c^{2}=\mathcal{O}(1)$ (here $M$ and $R$ are the characteristic mass and length scale of the phenomenon under consideration), is mandatory \cite{Will}.

This suggests that one might look for a theory that can provide a proper, more refined treatment in order to avoid singularities, at least in the paradigmatic situations arising in the strong fields mentioned above. Unfortunately, the fact that the singularity theorems use Einstein's equations only in a very weak sense --essentially only to conclude that gravitation is attractive-- rather suggest that the construction of a new theory without singular behavior, may constitute a difficult task.

Bearing this in mind, and not long ago \cite{Nos1},\cite{Nos2}, we exposed how to construct a gravitational action following the same guiding principles used many decades ago by Born and Infeld (BI) in the context of electrodynamics \cite{BI1}, \cite{BI2}. BI-like structures for the gravitational field were considered earlier in the literature. Historically they appear throughout two generations, the first, inaugurated by \cite{deser} and followed by other articles along the same line of research (see., e.g.,  Refs. \cite{Fein}-\cite{Cagri6}). In all these constructions, the gravitational action is obtained by combining higher order invariants built from the curvature in a Riemannian context, namely,

\begin{equation}
I_{\mathbf{BI(1)}}= \int d^{4}x\Big[\sqrt{|g_{\mu
\nu}+a\,R_{\mu\nu}+b\,X_{\mu\nu}|}-\sqrt{|g_{\mu\nu}|}\Big],
\label{acciondeser}
\end{equation}
where $|...|$ stands for the absolute value of the determinant, and $a,b$ are coupling constants. In (\ref{acciondeser}) we have separated the linear Ricci term $R_{\mu\nu}$ from the quadratic or higher order terms in the curvature contained formally in $X_{\mu\nu}$. The presence of all these curvature terms under the square root in expression (\ref{acciondeser}), is responsible for the fourth-order character of the field equations for the metric field $g_{\mu\nu}$. This fact complicates enormously the obtention of deformed exact solutions, i.e., solutions not present in GR and capable of shedding some light on the singularity problem.

In turn, in Ref. \cite{Max} another BI-like scheme based on a Palatini approach (actually called \emph{Eddington-Born-Infeld} gravity), was considered and first thoroughly studied mostly in cosmological environments, where a number of exact solutions without the big bang singularity were found \cite{Max2}-\cite{Avelino3}. The action in this case reads
\begin{equation}
I_{\mathbf{BI(2)}}= \int d^{4}x\Big[\sqrt{|g_{\mu
\nu}+a\,R_{\mu\nu}(\Gamma)|}-\sqrt{|g_{\mu\nu}|}\Big],
\label{accionmax}
\end{equation}
where $R_{\mu\nu}(\Gamma)$ represents the symmetric part of the Ricci tensor built with the connection $\Gamma$, which is taken as an independent field. In this second generation of BI theories, and due to the independent role played by the metric and the connection, second-order motion equations are obtained, even though they are different than Einstein's equations only when matter sources are present. This unfortunate fact eliminates the possibility of obtaining regular black hole states in pure vacuum. Very recent studies concerning stellar models and charged black holes within this framework can be found in Refs. \cite{EBI1}-\cite{EBI4}.

In what follows, we shall discuss the emergence of a geodesically complete bouncing cosmological solution in the context of the theory presented in \cite{Nos1} and \cite{Nos2}, which, unlike the second generation of BI gravitational theories just mentioned, is also able to deform vacuum general relativistic solutions. The spacetime in consideration represents an exact solution of the motion equations, which are second-order differential equations for the vielbein field $e^{a}(x)$ (unlike the ones coming from the first generation referenced above). The manuscript is organized in such a way that we briefly review the BI construction in section \ref{BI} below. Afterwards, in section \ref{Bbounce}, we obtain the solution and present a detailed discussion of its geometrical properties. Finally, we comment on its nature and discuss the regularity properties underlying the manifold thus obtained in section \ref{FC}.

Throughout the paper, we will adopt the signature $(+,-,-,...)$, and, as usual, Latin indexes $a:0,1,...$ refers to tangent-space objects while Greek $\mu:0,1,...$ allude to spacetime components.

\section{Born-Infeld gravity}\label{BI}

Following references \cite{Nos1} and \cite{Nos2}, we will assume that the dynamic of the gravitational field is described by the action in $D$ spacetime dimensions

\begin{equation}
I_{\mathbf{BIG}}=\frac{\lambda}{16 \pi G} \int d^{D}x\Big[\sqrt{|g_{\mu
\nu}+2\lambda ^{-1}F_{\mu \nu }|}-\sqrt{|g_{\mu \nu }|}\Big],
\label{acciondetelectro}
\end{equation}
where the tensor $F_{\mu \nu }$, the agent encoding the gravitational degrees of freedom, will be defined in brief (see eq. (\ref{tensorF}) below).

The action thus constructed provides an alternative dynamical behavior in the high energy regime, i.e, in situations where $\lambda ^{-1}F_{\mu \nu }=\mathcal{O}(1)$, where $\lambda$ is the Born-Infeld constant. In what follows, I shall briefly review the guiding principles leading to (\ref{acciondetelectro}). For details, the reader can consult the references just mentioned.

In order for (\ref{acciondetelectro}) to be a reasonable candidate for describing the gravitational field at length scales of order $\ell^{2}=\lambda ^{-1}$, we must ensure that the theory actually reduces to General Relativity in the low field limit. If we factor out $\sqrt{|g_{\mu \nu }|}$ in (\ref{acciondetelectro}) and use
\begin{equation}
\sqrt{|\mathbb{I}+2\lambda ^{-1}\mathbb{F}|}=1+\lambda ^{-1} Tr(\mathbb{F})+\mathcal{O}(\lambda ^{-2}),
\label{destraz}
\end{equation}
we get the action describing the low field limit
\begin{equation}
I_{\downarrow}=\frac{1}{16 \pi G} \int d^{D}x\, \sqrt{|g_{\mu \nu }|}\,Tr(\mathbb{F}),
\label{accionbaja}
\end{equation}
where $\mathbb{I}$ is the identity, and $\mathbb{F}\equiv F^{\nu}_{\mu}$. To elucidate the nature of the tensor $\mathbb{F}$ and its relation with the scalar curvature $R$ characterizing the Hilbert-Einstein action, we will require that the equations of motion for the fields responsible of the spacetime dynamics be of second order. Instead of being the metric tensor $\mathfrak{g}(x)$, we shall demand that the fundamental agent encoding the gravitational degrees of freedom consist of a set of $D$ 1-forms $\{e^{a}(x)\}$. A sufficient (though not necessary) condition for having second-order field equations is that the action (\ref{acciondetelectro}) includes up to first derivatives of $e^{a}(x)$, which means that $\mathbb{F}$ itself should be made up from this field and its first derivatives. This prescription, in view of the equivalence between (\ref{accionbaja}) and GR, poses what would appear at first sight to be an insurmountable problem, because $R$ contains second derivatives of the metric field. This problem disappears if we turn the attention to the absolute parallelism (or teleparallel) formulation of GR.

According to this point of view General Relativity can be formulated in a spacetime possessing
absolute parallelism. This approach is usually known as the
teleparallel equivalent of General Relativity TEGR
\cite{Hehl,Hehl2}, and it relies on the existence of a set
$\{e^{a}(x)\}$ of $D$ 1-forms that turn out to be autoparallel
for the Weitzenb\"{o}ck connection
\begin{equation}
\Gamma _{\mu \nu }^{\lambda
}=e_{a}^{\lambda }\,\partial _{\nu }e_{\mu }^{a},
\label{conexwei}
\end{equation}
where $e_{a}^{\lambda}$ refers to the inverse matrix of $e_{\mu }^{a}$. This connection
is curvature free, and it is compatible with the metric $\mathfrak{g}(x)=\eta _{ab}\
e^{a}(x)e^{b}(x)$, in the sense that the Weitzenb\"{o}ck covariant derivative of the metric vanishes. However, despite the fact that the curvature tensor associated to the Weitzenb\"{o}ck connection is null, the latter gives rise to a non-null torsion, which in the present context is simply\footnote{In general, the torsion 2-form is $T^{a}=De^{a}\ \doteq \ de^{a}+\mathbf{%
\omega}_{\ b}^{a}\wedge e^{b}$, where $\mathbf{%
\omega}_{\ b}^{a}$ the spin connection. But TEGR fixes $\mathbf{%
\omega}_{\ b}^{a}=0$. For details, the reader is invited to consult Ref. \cite{Nos3}.}
 \begin{equation}
T^{a}=de^{a},
\label{torweit}
\end{equation}
which means $T_{\ \ \mu \nu }^{\rho
}=e_{a}^{\rho }\,(\partial _{\mu }e_{\nu }^{a}-\partial _{\nu
}e_{\mu }^{a})$ in spacetime components. This ingredient can be combined in quadratic pieces in order to obtain a very remarkable identity, namely

\begin{equation}
R[e^a]\ =-T +\ 2\
\mathrm{e}^{-1}\,(T^{\mu\, \, \rho}_{\ \,
\mu}\,\mathrm{e})_{,\,\rho}\ , \label{equiv}
\end{equation}
where $\mathrm{e}=\sqrt{|\mathfrak{g}|}$ is the determinant of the
matrix $e^{a}_{\,\,\mu}$, $R$ is the scalar curvature, and the invariant $T$ is
\begin{equation}\label{weitinvariant}
T=S_{\rho }^{\ \ \mu \nu }T^{\rho}_{\ \ \mu \nu}.
\end{equation}

In this last equation we have introduced the important tensor
\begin{equation}
S_{\rho }^{\ \ \mu \nu }=-\frac{1}{4}\,(T_{\ \ \ \rho }^{\mu \nu
}-T_{\ \ \ \rho }^{\nu \mu }-T_{\rho }^{\ \ \mu \nu
})+\frac{1}{2}(\delta _{\rho }^{\mu }\,T_{\ \ \ \theta }^{\theta
\nu }-\delta _{\rho }^{\nu }\,T_{\ \ \ \theta }^{\theta \mu }).
\label{tensorS}
\end{equation}

Eq. (\ref{equiv}) is the central point in the equivalence between GR and TEGR, because it states that the scalar curvature constructed from the Levi-Civita connection can be explicitly viewed as a purely torsional object $T$ plus a total derivative. This peculiar invariant (usually known as the Weitzenb\"{o}ck invariant), is quadratic in the torsion tensor and, thus, is made up of the vielbein and its first derivatives alone.

Coming back to eq. (\ref{accionbaja}), we realize that in order to obtain the proper low energy limit given by Einstein's theory, we need to demand $Tr(\mathbb{F})=T$. This establishes the components of $\mathbb{F}$ according to\footnote{An antisymmetric second-rank tensor containing up to first derivatives of $e^{a}$ could be added to (\ref{tensorF}) without ruining the low energy limit of the theory. We shall not consider such a term in this work.}

\begin{equation}
F_{\mu \nu }=\alpha \,F^{(1)}_{\mu \nu}+\beta \,F^{(2)}_{\mu \nu}+\gamma \,F^{(3)}_{\mu \nu},  \label{tensorF}
\end{equation}%
where $\alpha ,\beta ,\gamma $ are dimensionless constants such as $\alpha
+\beta +D\, \gamma =1$ (hence, ensuring that $Tr(\mathbb{F})=T$), and the tensors $F^{(i)}_{\mu \nu}$ are defined by means of
\begin{equation}
F^{(1)}_{\mu \nu}=S_{\mu }^{\;\;\lambda \rho }T_{\nu \lambda \rho}\,,\,\,\,F^{(2)}_{\mu \nu}=S_{\lambda \mu }^{\;\;\;\rho }T_{\,\,\,\,\,\nu \rho }^{\lambda
}\,,\,\,\,F^{(3)}_{\mu \nu}=\,g_{\mu \nu }\,T\,. \label{tensorFies}
\end{equation}%

Then, we conclude by using the fundamental tensor (\ref{tensorF}) that the gravitational action (\ref{acciondetelectro}) reduces to GR in the low field limit. Additionally, due to the fact that it is constructed upon $e^{a}$ and its first derivatives alone, the motion equations coming from it are of second order.

What it makes action (\ref{acciondetelectro}) more interesting, is that it seems to incorporate a systematic treatment in order to avoid singularities. This constitutes the original idea behind the structures like (\ref{acciondetelectro}), whose spirit was fully understood first in the electromagnetic context by Born and Infeld \cite{BI1}, \cite{BI2}. Pictorially, the success of the determinantal structure concerning the singularity problem can be backtracked to the following simple fact: the tensor $\mathbb{F}$, by means of its symmetric part, allows us to think of a ``new'' metric tensor $\tilde{\mathfrak{g}}_{\mu\nu}(e^{a},e^{a}_{\,\,,\mu})$
\begin{equation}
\tilde{\mathfrak{g}}=\mathfrak{g}+2\lambda^{-1}F_{[\mu\nu]}\label{newmetric}
\end{equation}
Hopefully, this new metric can prevent the emergence of a singular state by moving geodesics away in a sort of repulsive regime at scales $\ell^{2}=\lambda ^{-1}$, thus evading the singularity theorems at such scales. Even though this repulsive high energy regime might not be a general feature of the theory \cite{DAS},\cite{Mariam}, it was shown that it actually exists under certain circumstances \cite{Nos1},\cite{Nos2}. In the next section we shall provide another example of it.

\section{A primordial Brusque Bounce}\label{Bbounce}

Historically, FRW cosmological spacetimes have been paradigmatic examples of singular behavior in GR. The big bang singularity, much like the $r=0$ Schwarzschild singularity, is an archetypical \emph{strong curvature singularity}, as defined in \cite{Tipler} and \cite{Tipler2} (see also \cite{Ellis} for further comments on this concept). Undoubtedly, this kind of singular state constitutes the most explicit and harmful situation envisioned within the general relativistic context.

Now we want to show how a radically different description of the very early stages of the Universe emerges out as a consequence of action (\ref{acciondetelectro}). For this reason we have to prescribe a frame field $e^{a}$ for spatially flat isotropic and homogeneous FRW manifolds first. An appropriated parallelization of these spacetimes is provided by the frame \cite{Nos4}
\begin{equation}
e_{\mu }^{a}=\emph{diag}(1,\, a(t),\, a(t),\, ...),\ \ \ \ \ \
e=a^{D-1}. \label{frame}
\end{equation}
This frame not only leads to the metric tensor
\begin{equation}
\mathfrak{g}=\emph{diag}(1,-a(t)^2,-a(t)^2,...), \label{metricafrw}
\end{equation}
but also constitutes a globally well-defined basis for the cotangent space $\mathcal{T}^{*}(\mathcal{M})$, being $\mathcal{M}=(\mathbb{R}^{4},\mathfrak{g})$.
In order to describe the cosmic evolution, we will assume a perfect fluid with energy density $\rho$ and pressure $p$ as source of the motion equations. We have then,
\begin{equation}
T^{\mu\nu}=(\rho+p)V^{\mu}V^{\nu}+p\, g^{\mu\nu},\label{tenene}
\end{equation}%
where $V^{\mu}$ is the tangent vector to the congruence of causal curves defining the flow lines. Additionally we shall suppose that such a perfect fluid is isentropic, so that we will have $p=\omega\rho$, with$\omega$ the barotropic index. The energy-momentum tensor (\ref{tenene}) adopts a very simple form in the comoving frame, where it simply reads $T^{\mu}_{\,\,\nu}=diag(\rho,-\omega\rho,-\omega\rho,... )$.

The motion equations are obtained by varying the action (\ref{acciondetelectro}) with respect to the vielbein components $e^{a}_{\mu}$. In the following, we will be interested in the case $D=4$. For the tetrad (\ref{frame}) we can easily check that the only non-null components of $T_{\mu\nu\rho}$ and $S_{\mu\nu\rho}$ (see eqs. (\ref{torweit}) and (\ref{tensorS}) respectively), are
\begin{eqnarray}  \label{sandtes}
&&S_{\mu 0\mu }=-S_{\mu \mu 0}=-\ a(t)\ \dot{a}%
(t),  \notag \\
&&T_{\mu 0\mu }=-T_{\mu \mu 0}=a(t)\ \dot{a}(t), \ \ \
\ \ \ \ \ \mu \neq 0,
\end{eqnarray}
so, the  Weitzenb\"{o}ck (\ref{weitinvariant}) results in $T=-6H^{2}$, where $H=\dot{a}/a$ is the Hubble rate, and the dots refer to derivatives with respect to the proper time $t$.
With these components in hand, we can evaluate the three pieces of $F_{\mu\nu}$ according to eqs. (\ref{tensorFies}). These are

\begin{eqnarray}  \label{compfies}
&&F^{(1)}_{\mu \nu}=diag(0,2\,\dot{a}^{2},2\,\dot{a}^{2},2\,\dot{a}^{2}), \notag \\
&&F^{(2)}_{\mu \nu}=diag(-3H^{2},\dot{a}^{2},\dot{a}^{2},\dot{a}^{2}), \notag \\
&&F^{(3)}_{\mu \nu}=diag(-6H^{2},6\,\dot{a}^{2},6\,\dot{a}^{2},6\,\dot{a}^{2}).
\end{eqnarray}

The constraint equation (the one coming from the variation with respect to $e^{0}_{\,0}$), reads

\begin{equation}
\frac{\sqrt{1-B H^2}}{\sqrt{1-A H^2}}[1+2B H^2-3 AB H^4]-1=\frac{16 \pi G }{\lambda}\rho, \label{valink=0gen}
\end{equation}%
where
\begin{equation}
A=6(\beta+2\gamma)/\lambda,\,\,\,\,\,\,B=2(2\alpha+\beta+6\gamma)/\lambda, \label{condiciones}
\end{equation}
As usual, the conserved character of $T^{\mu\nu}$ gives rise to

\begin{equation}
\frac{d}{dt}(\rho\, a^{3})=-p\frac{d}{dt}(a^{3})\,\,\rightarrow \,\,\dot{\rho}+3(\rho+p)H=0 . \label{conflui}
\end{equation}
For the isentropic case under consideration, this last equation acquires the familiar form
\begin{equation}
\rho(t)=\rho_{0} \Big(\frac{a_{0}}{a(t)}\Big)^{3(1+\omega)},\label{eqcon}
\end{equation}
where $a_{0}$ and $\rho_{0}$ are two integration constants.

From now on, we will focus on the important case where $A=B$ in (\ref{valink=0gen}), which is susceptible to an analytical treatment\footnote{Different choices of $\alpha$, $\beta$ and $\gamma$ in a cosmological context were considered earlier in Ref. \cite{Nos2}}. Due to $(\ref{condiciones})$, this implies $\alpha=\beta$ and a free $\gamma$ parameter in the action, which can be reabsorbed in $\lambda$. Redefining $\lambda\rightsquigarrow
(2A)^{-1}\lambda$, the equation (\ref{valink=0gen}) reads in this case

\begin{equation}
6H^2 \Big(1-\frac{9H^2}{2\lambda}\Big)=16\pi G\ \rho.\label{valink=0part}
\end{equation}

From this expression, the Hubble rate is easily obtained as

\begin{equation}\label{ramasH}
H^2=\frac{\lambda}{9}\Big(1\pm\sqrt{1-3\mathtt{y}}\Big),
\end{equation}
where we have defined the non-dimensional variable
\begin{equation}
\mathtt{y}=\frac{16 \pi G}{\lambda}\rho(t).\label{vary}
\end{equation}
From this definition it follows that
\begin{equation}
\dot{\mathtt{y}}=-3(1+\omega)\, H\,\mathtt{y}.\label{vary}
\end{equation}

The branch with positive sign in Eq. (\ref{ramasH}) (referred hereafter as the \emph{positive branch}), reveals itself as a purely high energy state, disconnected from the GR limit. This is so because it leads to a maximum Hubble rate as the density goes to zero, or equivalently, when $a_{0}/a(t)\rightarrow 0$. It is clear then, that this configuration must be excluded from the physically admissible solution space.

The \emph{negative branch} (i.e., the one with negative sign in Eq. (\ref{ramasH})), behaves differently according to the sign of the BI constant $\lambda$. Note that even when $\lambda<0$ the equation is well defined. However, we will show in short that this situation leads us to no regularization process at all, because the Hubble rate diverges as the scale factor goes to zero just as in GR. In turn, if $\lambda>0$, the negative branch shows that the Hubble parameter in terms of the variable $\mathtt{y}$ reaches a maximum value given by $H_{max}^2=\lambda/9$, which corresponds to a maximum energy density $\rho_{max}=(48 \pi G)^{-1}\lambda$. This is the type of cosmic evolution that we want to discuss thoroughly in the next paragraphs.

Whatever the barotropic index is (excepting the case $\omega=-1$, which can be worked out by introducing a cosmological constant term in the action), the variable $\mathtt{y}$ allows us to integrate the equation (\ref{ramasH}) in closed form. As a matter of fact, we have from (\ref{ramasH}) and (\ref{vary}) that

\begin{equation*}
\dot{\mathtt{y}}=\pm\mathcal{A}\,\mathtt{y}\,\sqrt{1-\sqrt{1\pm3\mathtt{y}}},
\end{equation*}

where $\mathcal{A}=(1+\omega)\sqrt{|\lambda|}$ is a non-null constant. Note that the $\pm$ on the right-hand side of this equation comes from taking square root in (\ref{ramasH}), and should not be confused with the ones coming from the different branches, which appear in the radicand. This equation can be straightforwardly integrated, leading to the following expressions according to the sign of $\lambda$:

\begin{equation}\label{intramasH1}
\pm\tilde{\mathcal{A}}\, t\pm
c=(f^{-}(\mathtt{y}))^{-1}- arctan(f^{-}(\mathtt{y})),\,\,\,\,\lambda<0,
\end{equation}

\begin{equation}\label{intramasH2}
\pm\tilde{\mathcal{A}}\, t\pm
c=(f^{+}(\mathtt{y}))^{-1}+ arctanh(f^{+}(\mathtt{y})),\,\,\,\,\lambda>0,
\end{equation}
where $c$ is an integration constant.

In these equations we have defined $\tilde{\mathcal{A}}=\mathcal{A}/\sqrt{2}$, and the functions $f^{+/-}(\mathtt{y})$ are such that

\begin{eqnarray}\label{defy}
f^{-}(\mathtt{y})&=&\Big(\frac{-1+\sqrt{1+3\mathtt{y}}}{2}\Big)^{1/2}\\
f^{+}(\mathtt{y})&=&\Big(\frac{1-\sqrt{1-3\mathtt{y}}}{2}\Big)^{1/2}.
\end{eqnarray}
 Starting from the exact expressions (\ref{intramasH1}) and (\ref{intramasH2}), we can characterize the scale factors as functions of the proper time $t$. Naturally, irrespective of the sign of $\lambda$ we have the GR limit when $\mathtt{y}\rightarrow0$, i.e., when $|\lambda|\rightarrow\infty$. In this limit, equations (\ref{intramasH1}) and (\ref{intramasH2}) become

\begin{equation}\label{GRlimit}
\frac{a(t)}{a_{0}}= \Big(\frac{3}{2} H_{0}(1+\omega)\,t\Big)^{2/3(1+\omega)},
\end{equation}
where $H^{2}_{0}=8\pi G \rho_{0}/3$ according to the Friedmann equation. This last equation describes the dynamics of the scale factor in GR.

From now on, and due to the fact that we are interested in the very early stages of the cosmic evolution, we shall focus on a radiation filled Universe ($\omega=1/3$). Similar results are obtained for other values of the barotropic index, provided $\omega\neq-1$.

For the case of $\lambda<0$, an expansion of (\ref{intramasH1}) in the small quantity $a(t)/a_{0}$ give us at first order

\begin{equation}\label{smallneg}
\frac{a(t)}{a_{0}}\propto \sqrt{\lambda}\,\, t,\,\,\,\,\,\,\,\,H\propto\,t^{-1}.
\end{equation}
From the point of view of singularities, the case $\lambda<0$ and its underlying dynamics given by (\ref{smallneg}) are as unsatisfactory as they are in GR. The resulting spacetime is geodesically incomplete, and the energy density and pressure become unbounded as $t\rightarrow 0^{+}$. In other words, the spacetime does not admit an extension in $t=0$ and any pair of events $(t_{1},t_{2})$ with $t_{1}<0$ and $t_{2}>0$ cannot be connected by any causal geodesic.

A quite different dynamic is obtained when one considers $\lambda>0$. Around $\mathtt{y}=1/3$, Eq. (\ref{intramasH2}) can be approximated by

\begin{equation}\label{smallpos}
\Big(\frac{a(t)}{a_{0}}\Big)^{4}=\frac{48 \pi G \rho_{0}}{\lambda(1\pm 4\sqrt{\lambda}\,t)}+\mathcal{O}(\lambda\,t^{2}),
\end{equation}
whereas the Hubble rate is
\begin{equation}\label{smallpos1}
H(t)=\mp\frac{\sqrt{\lambda}}{(1\pm4\sqrt{\lambda}\,t)}+\mathcal{O}(\lambda\,t^{2}).
\end{equation}
It is important to note that in these equations (and in the following throughout the paper) the minus sign corresponds to $t>0$, and the plus to $t<0$ in the term $(1\pm4\sqrt{\lambda}\,t)$. All the remaining signs in the expressions must preserve the right order. For instance, positive times require the plus sign on the right hand side of Eq. (\ref{smallpos1}).

The event $t=0$ is called a \emph{brusque bounce} because there exists a minimum scale factor
\begin{equation}\label{minscale}
\frac{a_{min}}{a_{0}}=\Big(\frac{48 \pi G \rho_{0}}{\lambda}\Big)^{1/4},
\end{equation}
for which $H^{2}\neq0$. Actually,  $H$ is not defined at $t=0$ because $\lim_{t\rightarrow 0^{\pm}}H(t)=\pm\sqrt{\lambda}$, even though $H^{2}$ is perfectly behaved there with value $H^{2}(0)=\lambda$, see Eq. (\ref{smallpos1}). This pathology is not as dangerous as it might seem at first glance, because it does not jeopardize the $\mathcal{C}^{1}$ character of causal geodesics at $t=0$. This can be explicitly checked by means of the geodesic equation, or even easily, by taking advantage of the six conserved quantities associated to the isotropy and homogeneity of FRW spacetimes.  It is not hard to see that in standard spherical coordinates $(r,\theta,\phi)$, the tangent vector of the geodesics is described by \cite{Lazkoz}
\begin{eqnarray}
(\partial_{\tau}t)^{2}&=&\delta+ \frac{P^{2}}{a(t)}\notag\\
\partial_{\tau}r&=&\frac{P_{1}\,cos\phi+P_{2}\,sin\phi}{a(t)}\notag\\
\partial_{\tau}\phi&=&\frac{L_{3}}{a(t)r^{2}},\label{geo}
\end{eqnarray}
where $\delta$ is 0 or 1 according to the null or timelike character of the geodesic respectively, and $\tau$ is an affine parameter. Note that, due to spherical symmetry, every geodesic may be constrained to lay in the hypersurface $\theta=\pi/2$, and $L_{1}=L_{2}=P_{3}=0$ by an appropriated coordinate change. In Eq. (\ref{geo}), $P^{2}=P_{1}^{^{2}}+P_{2}^{^{2}}+P_{3}^{^{2}}$ is the total linear momentum and $L^{2}=L_{1}^{^{2}}+L_{2}^{^{2}}+L_{3}^{^{2}}$, the total angular momentum (being $P_{i}$ and $L_{i}$, $i:1,2,3$ six constants of motion associated to the six-dimensional group of isometries characteristic of FRW models). So, due to the fact that $a(t)\neq0$ for all $t$, the geodesics are $\mathcal{C}^{1}$ curves.

According to (\ref{geo}) the acceleration vector involves first derivatives of the scale factor, which are not defined in the bounce. This means that the Riemann tensor itself and the scalars constructed from it are not defined in $t=0$. Nonetheless, the behavior of the Riemann tensor as one approaches the bounce for either side of the time variable is not divergent. For instance, we can evaluate the scalar curvature for the obtained solution in the vicinity of the bounce. In order to do this we have to compute $R=6(2H^{2}+\dot{H})$ for the scale factor (\ref{smallpos}). It results
\begin{equation}\label{Curvsca}
R=\frac{12\lambda\,(1\pm2)}{(1\pm4\sqrt{\lambda}\,t)^{2}}+\mathcal{O}(\lambda\,t^{2}).
\end{equation}
Other invariants suffer from this indefiniteness at the bounce too. Examples of these are the quadratic scalars such as $R^{2}=R^{\mu\nu}R_{\mu\nu}$ and $K=R^{\mu}_{\,\,\,\nu\lambda\rho}R_{\mu}^{\,\,\,\nu\lambda\rho}$, because they not only involve products of the form $H^{4}$ and $\dot{H}^{2}$ (which are well defined), but also a term $H^{2}\dot{H}$ (which is not).

One may wonder, thus, what the effect of the bounce is not only on point particles, but also on extended finite objects. In order to examine this issue, we must note that the event $p_{0}$ given by $t=0$ is not a strong curvature singularity, as defined by Tipler in \cite{Tipler}. We can see this explicitly by examining the expression

\begin{equation}\label{tipler}
\mathbb{T}=\lim_{t\rightarrow 0}\int R_{\mu\nu}K^{\mu} K^{\nu}\, dt,
\end{equation}
where $K^{\mu}$ is the tangent vector of every null geodesic generator $\lambda(t)$ that intersects the point $p_{0}$ at affine parameter value 0. It was shown that $\mathbb{T}<\infty$ captures the physical requirement that an extended finite object is not crushed to zero volume by the effect of tidal forces. Condition $\mathbb{T}<\infty$ can be fulfilled if for every interval $(0,t_{1})$ there is an affine parameter $t_{2}\in(0,t_{1})$ such that
\begin{equation}\label{tipler2}
R_{\mu\nu}K^{\mu} K^{\nu}\mid_{t=t_{2}}<t_{2}^{-q},
\end{equation}
for any fixed $q<1$ \cite{Tipler}. We proceed now to prove that (\ref{tipler2}) (and then $\mathbb{T}<\infty$), actually holds.

The Ricci tensor for spatially flat FRW cosmologies reads
\begin{equation}\label{ricci}
R_{00}=3(H^{2}+\dot{H})\,\,\,\,\,\hat{R}_{\mu\nu}=(3H^{2}+\dot{H})\hat{g}_{\mu\nu},
\end{equation}
 where $\hat{R}_{\mu\nu}$ and $\hat{g}_{\mu\nu}$ refer to purely spatial Ricci and metric tensors respectively. In a neighborhood of $t=0$, we can make use of expressions (\ref{smallpos}) and (\ref{smallpos1}), and by means of the velocity vectors (\ref{geo}), we can compute the left-hand side of Eq. (\ref{tipler2}). For $t>0$ it reads
\begin{equation}\label{tipler3}
R_{\mu\nu}K^{\mu} K^{\nu}=\frac{C_{1}}{(1-4\sqrt{\lambda}\,t)^{2}}-\frac{C_{2}}{(1-4\sqrt{\lambda}\,t)^{7/4}},
\end{equation}
where $C_{1}$ and $C_{2}$ are positive functions of the constants of motion and of the spatial coordinates, but not functions of the proper time. We clearly have then
\begin{equation}\label{tipler4}
R_{\mu\nu}K^{\mu} K^{\nu}<\frac{C_{1}}{(1-4\sqrt{\lambda}\,t)^{2}}.
\end{equation}
We immediately note that
\begin{equation}\label{tipler4}
\frac{C_{1}}{(1-4\sqrt{\lambda}\,t)^{2}}<t^{-q},
\end{equation}
for any $t$ sufficiently close to zero and any $q<1$, as a consequence of the divergent character of $t^{-q}$ as $t\rightarrow 0^{+}$. This establishes (\ref{tipler2}), and the fact that the brusque bounce does not crush extended finite objects to zero volume by the effect of tidal forces.

This means that the spacetime, regarded as the pair $(\mathcal{T}^{*}(\mathcal{M}), e^{a})$, admits a $\mathcal{C}^{0}$ local extension at $p_{0}$. This consists in taking the two signs in (\ref{intramasH2}) and gluing together the two scale factors at the event $p_{0}$, i.e., $\mathcal{T}^{*}(\mathcal{M})=\mathcal{T}^{*}(\mathcal{M}_{1})\cup\mathcal{T}^{*}(\mathcal{M}_{2})$, where $\mathcal{M}_{1}=(\mathbb{R}^{4},\mathfrak{g}, t\leq0)$, and $\mathcal{M}_{2}=(\mathbb{R}^{4},\mathfrak{g}, t\geq0)$, with $\mathfrak{g}=dt^{2}-a^{2}(t)(dx^{2}+dy^{2}+dz^{2})$. This procedure, of course, will encompass the right election of the integration constant appearing in (\ref{intramasH2}).

Finally, the maximal extension so constructed is globally hyperbolic. In terms of the conformal time
\begin{equation}\label{conformaltime}
\tilde{\tau}=\int a^{-1}(t)\,dt,
\end{equation}
which is well defined for $-\infty<t<\infty$ by virtue of the everywhere non-null (and $\mathcal{C}^{0}$) character of $a(t)$, the extension is conformal to Minkowski spacetime. Any hypersurface of constant time ($t=0$ among them), constitutes a Cauchy surface.

\begin{figure}[ht]
\centering
\includegraphics[scale=.75]{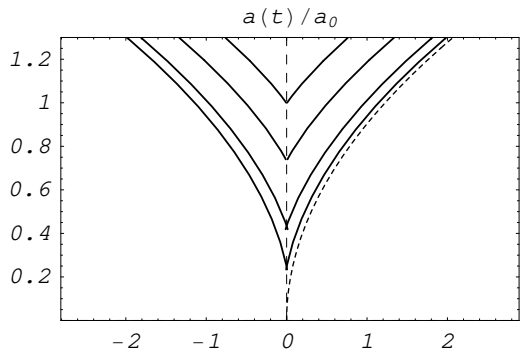} \caption[]{The scale factor that emerges from Eq. (\ref{intramasH2}), for four different values of the BI constant. From top to bottom in the solid curves: $\lambda=3,10,10^{2},10^{3}$. The dashed curve represents the $\omega=1/3$ GR Universe.}\label{bouncing}
\end{figure}

The scale factor thus obtained from Eq. (\ref{intramasH2}) is visualized in Figure \ref{bouncing}, where curves with four different values of the Born-Infeld constant $\lambda$ are depicted. From top to bottom, the solid lines represent the values $\lambda=3,10,10^{2},10^{3}$, where for simplicity we have taken $16 \pi G \rho_{0}=1$. Furthermore, we have included in Figure \ref{bouncing} the corresponding GR curve for a radiation-dominated Universe (dashed line), namely $a(t)/a_{0}=\sqrt{2H_{0}\, t}$ (see Eq. (\ref{GRlimit})).

\section{Final remarks}\label{FC}

In Einstein´s Theory of General Relativity no physical process exists in order to avoid the inevitable fact that, far back in time, the energy density and pressure of the matter fields were infinite as a consequence of the vanishing of the scale factor at the Big Bang. We have shown throughout the preceding paragraphs, that Born-Infeld gravity offers a quite different description of the very early stages of the cosmic evolution, for it was shown in this article that the dynamical equations (\ref{valink=0gen}) for $A=B$ lead to a bounce of the scale factor where the Universe possesses a minimum size given in (\ref{minscale}) for a radiation-like Universe. In general, for $\omega\neq-1$ it is easy to show that the minimum 3-volume associated to the bounce is

\begin{equation}\label{parcela}
\Big(\frac{a_{min}}{a_{0}}\Big)^3=\Big(\frac{48 \pi G
\rho_{0}}{\lambda}\Big)^{1/(1+\omega)}.
\end{equation}

Throughout this work we have advocate for not considering the event $t=0$ as a singularity, at least, not in any of the widely accepted senses. This viewpoint is supported by at least two strong arguments:

a) On the one hand, we stressed in the last section that the geodesics are well behaved at the bounce, in the sense that they are $\mathcal{C}^{1}$ curves there. Actually, the spacetime is geodesically complete, and all the causal geodesics are of class $\mathcal{C}^{\infty}$, except at the bounce itself. Moreover, the Tipler condition $\mathbb{T}<\infty$ assures us not only that pointlike particles traveling along causal geodesics do not experience any kind of singular behavior, but that extended objects of finite size do not crush to zero volume in passing the bounce. Nonetheless, these two minimum conditions are also satisfied, for instance, by the so-called sudden singularities \cite{Barrow1}-\cite{Barrow4}.

b) On the other hand, in the brusque bounce both, the energy density and the pressure of the matter fields are perfectly well behaved and they result finite because we assume throughout the analysis a very simple equation of state of the form $p=\omega\rho$. Actually, we have from (\ref{eqcon}) and (\ref{smallpos}) that sufficiently close to the bounce the energy density (for $\omega=1/3$) scales as

\begin{equation}\label{enden}
\rho(t)=\frac{\lambda}{48 \pi G}(1 \pm 4\sqrt{\lambda}\,t),
\end{equation}
so a maximum energy density $\rho_{max}=\lambda/48 \pi G$, and a maximum pressure $p_{max}=\lambda/144 \pi G$ are reached at the very bounce. This is not the case, for instance, in the sudden singularities, where the energy density is finite but not the pressure, but it is the case for the type IV singularities. In the latter the matter fields are finite, but divergences in higher order derivatives of the Hubble rate occur (even though $H$ and $\dot{H}$ are finite) \cite{typeIV1}.

However, a word of caution should be said about this optimistic point of view concerning the bounce as a regular event. If we insist on viewing the metric tensor as the dynamical field that encodes the geometrical properties of the spacetime, the very fact that the Riemann curvature tensor is not defined at the bounce --as well as the whole repertoire of curvature invariants constructed from it-- is certainly unsatisfactory. As long as we profess this Riemannian philosophy, it is inevitable to think about the bounce as a singularity.

But the theory here exposed is not based on Riemannian concepts, even though the metric emerges as an agent that confers orthonormality to the vielbein field $e^{a}$, which embodies the spacetime geometrical structure by means of a parallelization process. This global basis of the cotangent bundle constitutes a preferred reference frame that can be used to define the space structure, in the sense that we can define a given spacetime as the pair $(\mathcal{T}^{*}(\mathcal{M}_{D}),e^a(x))$ instead of $(\mathcal{M}_{D},g_{\mu\nu}(x))$. This preferred frame is, nonetheless, not unique. In ref. \cite{Nos3} it was established the existence of a remnant group of Lorentz transformations in the so-called $f(T)$ gravity, which constitutes a modified scheme for the description of the gravitational field based upon the same geometrical structure of the theory here considered. Preliminary results show that this is actually the case also in the present context, which means that for every spacetime $(\mathcal{T}^{*}(\mathcal{M}_{D}),e^a(x))$ there exists a subgroup $\mathcal{A}(e^a)$ of the Lorentz group such that $e^{a\,'}(x)=\Lambda^{a\,'}_{a}e^a(x)$ describes the same spacetime, for $\Lambda^{a\,'}_{a}\in\mathcal{A}(e^a)$. In any case, it is clear that the motion equations of the theory under consideration determine the full tetrad components (up to transformations of the potential remnant group), and not just the metric tensor. To look for pathologies concerning the Riemann tensor and the scalars coming from it (i.e., coming from the metric and its first and second derivatives), seems manifestly misleading \footnote{However, the energy-momentum tensor of spinless matter couples to the metric $\mathfrak{g}=e^{a}e^{b}\eta_{ab}$ and not to the tetrad in the present formalism. In the case of FRW spacetimes, and by virtue of the exceedingly simple state equation $p=\omega\rho$, eq. (\ref{eqcon}) holds, so the matter field couple to the scale factor (which is defined everywhere), and not to its derivatives. Care should be taken in supporting this point of view under more general circumstances.}.

These comments tend to emphasize that, beyond Einstein's Theory of General Relativity, and in the context of extended theories of gravity formulated by purely torsional means, the criterion of a singularity based on unbounded large (or even undefined) values of scalar polynomials in the Riemann curvature, becomes strongly objectionable. One should pay attention to the tensors and scalars constructed in Weitzenb\"{o}ck spacetime instead. If we focus on invariants containing just first derivatives of the vielbein, we have that they behave as $H^{2n}$, with $n$ a positive integer. Precisely, we know from (\ref{sandtes}) that $T=-6\ H^{2}$, and with the help of (\ref{tensorF}) and (\ref{tensorFies}), we obtain

\begin{equation}\label{trazasvar}
Tr(\mathbb{F}^{n}) \propto H^{2n},\,\,\,\,\mathbb{F}^{n}\equiv F^{\mu}_{\,\,\nu1}F^{\nu1}_{\,\,\nu2}...F^{\nu(n-1)}_{\,\,\rho}.
\end{equation}
All these Weitzenb\"{o}ck scalars are well behaved throughout the whole cosmic evolution, because so it is $H^{2}$. The same is true for the action itself, which is nothing more than combinations of terms of the form (\ref{trazasvar}).

The results we have obtained support the idea that the Universe did not begin a finite time ago. This is not actually a new idea, for in many alternative approaches to the description of the very early Universe we find similar outcomes (see, for instance \cite{Novello}, and references therein contained). Remarkably, this conclusion is obtained here as a consequence of assuming a dynamic for the matter fields as simple as $p=\omega\rho$ and barotropic indexes with a very diaphanous physical interpretation, such as $\omega=0,1/3$.



\emph{Acknowledgements}. The author wants to thanks J. Areta and A. Borrelli for a careful reading of the manuscript and for helpful suggestions about it. This work was supported by CONICET and Instituto Balseiro.


\begin{thebibliography}{99}


\bibitem{H-P} S. W. Hawking and R. Penrose, \emph{Proc. Roy. Soc. Lond} {\bf A314} (1970) 529.
\bibitem{H-E} S. W. Hawking and G. F. R. Ellis, \emph{The large scale structure of spacetime}, Cambridge University Press, Cambridge (1973).
\bibitem{Geroch} R. Geroch, \emph{Annals of Physics} {\bf 48} (1968) 526.
\bibitem{Joshi1} For a recent account, see P. Joshi, Spacetime singularities, in \emph{Springer handbook of spacetime}, A. Ashtekar and V. Petkov Eds. (2014) 409.
\bibitem{Clarke1} C. J. S. Clarke, \emph{Gen. Rel. Grav.} {\bf 6} (1975) 35.
\bibitem{Ellis} G. F. R. Ellis and B. G. Schmidt, \emph{Gen. Rel. Grav.} {\bf8} (1977) 915.
\bibitem{TCE} F. J .Tipler, C. J. S. Clarke and G. F. R. Ellis, \emph{Singularities and Horizons: A Review Article}, in General Relativity and Gravitation, A. Held, Ed. (Plenum Press, New York, 1980) 97.
\bibitem{Schmidt1} B. G. Schmidt, \emph{Gen. Rel. Grav.} {\bf 1} (1971) 269.
\bibitem{Schmidt2} B. G. Schmidt, \emph{Comm. Math. Phys.} {\bf 29} (1973) 49.
\bibitem{Szekeres} S. M. Scott and P. Szekeres, \emph{J. Geom. Phys.} {\bf 13} (1994) 223.
\bibitem{Barrow1} J. D. Barrow, \emph{Class. Quant. Grav.} {21} (2004) L79.
\bibitem{Barrow2} J. D. Barrow, \emph{Class. Quant. Grav.} {21} (2004) 5619.
\bibitem{Barrow3} J. D. Barrow and C. G. Tsagas, \emph{Class. Quant. Grav.} {22} (2005) 1563.
\bibitem{Barrow4} J. D. Barrow, S. Cotsakis and A. Tsokaros, \emph{Class. Quant. Grav.} {27} (2010) 165017.
\bibitem{Lazkoz} L. Fernández-Jambrina and R. Lazkoz, \emph{Phys. Rev.} {\bf D70} (2004) 121503.
\bibitem{Taylor} J. H. Taylor and J. M. Weisberg, \emph{ApJ} {\bf345} (1989) 434.
\bibitem{Ligo} B. P. Abbot et al, \emph{Phys. Rev. Lett.} {\bf 116} (2016) 061102.
\bibitem{Will} C. Will, \emph{Living. Rev. Relativity} {\bf17} (2014) 4.
\bibitem{Nos1} R. Ferraro and F. Fiorini, \emph{Phys. Lett.} {\bf B692} (2010) 206.
\bibitem{Nos2} F. Fiorini, \emph{Phys. Rev. Lett.} {\bf 111} (2013) 041104.
\bibitem{BI1} M. Born and L. Infeld, Proc. R. Soc. \textbf{A144} (1934) 425.
\bibitem{BI2} M. Born and L. Infeld, Proc. R. Soc. \textbf{A147} (1934) 522.
\bibitem{deser} S. Deser and G.W. Gibbons, Class. Quant. Grav. \textbf{15} (1998) L35.
\bibitem{Fein} J.A. Feingenbaum, Phys. Rev. \textbf{D58} (1998) 124023.
\bibitem{Fein2} J. A. Feingenbaum, P.O. Freund and M. Pigli, Phys. Rev. \textbf{D57} (1998) 4738.
\bibitem{Comelli} D. Comelli and A. Dolgov, JHEP \textbf{0411} (2004) 062.
\bibitem{Nieto}J.A. Nieto, Phys. Rev. \textbf{D70} (2004) 044042.
\bibitem{Com1} D. Comelli, Phys. Rev. \textbf{D72} (2005) 064018.
\bibitem{Vollick} D. N. Vollick, Phys. Rev. \textbf{D72} (2005) 084026.
\bibitem{Cagri1} I. Gullu, T. Cagri Sisman and B. Tekin, Class. Quant. Grav. \textbf{27} (2010) 162001.
\bibitem{Cagri2} I. Gullu, T. Cagri Sisman and B. Tekin, Phys. Rev. \textbf{D81} (2010) 104018.
\bibitem{Cagri3} I. Gullu, T. Cagri Sisman and B. Tekin, Phys. Rev. \textbf{D82} (2010) 024032.
\bibitem{Cagri4} I. Gullu, T. Cagri Sisman and B. Tekin, Phys. Rev. \textbf{D82} (2010) 124023.
\bibitem{Cagri5} I. Gullu, T. Cagri Sisman and B. Tekin, Phys. Rev. \textbf{D91} (2015) 044007.
\bibitem{Cagri6} I. Gullu, T. Cagri Sisman and B. Tekin, Phys. Rev. \textbf{D92} (2015) 104014.
\bibitem{Max} M. Ba\~{n}ados, Phys. Rev. \textbf{D77} (2008) 123534.
\bibitem{Max2} M. Ba\~{n}ados, P. G. Ferreira and C. Skordis, Phys. Rev. \textbf{D79} (2009) 063511.
\bibitem{Max3} M. Ba\~{n}ados and P. G. Ferreira, Phys. Rev. Lett. \textbf{105} (2010) 011101.
\bibitem{Max4} C. Escamilla-Rivera, M. Ba\~{n}ados and P. G. Ferreira, Phys. Rev. \textbf{D85} (2012) 087302.
\bibitem{Max5} J. H. C. Scargill, M. Ba\~{n}ados and P. G. Ferreira, Phys. Rev. \textbf{D86} (2012) 103533.
\bibitem{Terence1} P. Pani, V. Cardoso and T. Delsate, Phys. Rev. Lett. \textbf{107} (2011) 031101.
\bibitem{Terence2} T. Delsate and J. Steinhoff, Phys. Rev. Lett. \textbf{109} (2012) 021101.
\bibitem{Avelino1} P. P. Avelino and R. Z. Ferreira, Phys. Rev. \textbf{D86} (2012) 041501.
\bibitem{Avelino2} P. P. Avelino, Phys. Rev. \textbf{D85} (2012) 104053.
\bibitem{Avelino3} P. P. Avelino, JCAP 1211 (2012) 022.
\bibitem{EBI1} H. Sotani, Phys. Rev. \textbf{D89} (2014) 124037.
\bibitem{EBI2} H. Sotani and U. Miyamoto Phys. Rev. \textbf{D90} (2014) 124087.
\bibitem{EBI3} H. Sotani and U. Miyamoto Phys. Rev. \textbf{D91} (2015) 084020.
\bibitem{EBI4} H. Sotani and U. Miyamoto Phys. Rev. \textbf{D92} (2015) 044052. 	
\bibitem{Hehl} J. Nitsch and F.W. Hehl, Phys. Lett. \textbf{B90} (1979) 98.
\bibitem{Hehl2} F.W. Hehl, J. D. McCrea, E. W. Mielke and Y. Ne'eman, Phy. Rept. \textbf{258} (1995) 1.
\bibitem{Nos3} R. Ferraro and F. Fiorini, \emph{Phys. Rev.} {\bf D91} (2015) 064019.
\bibitem{DAS} S. Jana, \emph{Phys. Rev.} {\bf D90} (2014) 124007.
\bibitem{Mariam} M. Bouhmadi-Lopez, C-Y Chen and P. Chen, \emph{Phys. Rev.} {\bf D90} (2014) 123518.
\bibitem{Nos4} For a discussion on the problem of parallelization in theories with absolute parallelism see, for instance, F. Fiorini, P.~A. Gonz\'{a}lez and Y. V\'{a}squez, \emph{Phys. Rev.} \textbf{D89} (2014) 024028.
\bibitem{Tipler} F. J. Tipler, \emph{Phys. Lett.} {\bf A64} (1977) 8.
\bibitem{Tipler2} F. J. Tipler, \emph{Nature} {\bf 270} (1977) 500.
\bibitem{typeIV1} S. Nojiri, S. D. Odintsov and S. Tsujikawa, \emph{Phys. Rev.} {\bf D71} (2005) 063004.
\bibitem{Novello} M. Novello and S. E. Perez Bergliaffa, Phys. Rept. \textbf{463} (2008) 127.
\end{thebibliography}
\end{document}